\documentclass[conference]{IEEEtran}
\ifCLASSINFOpdf
  \usepackage[pdftex]{graphicx}
  \graphicspath{{figures/}}
  \DeclareGraphicsExtensions{.pdf,.jpeg,.png}
\else
\fi
\begin{document}
\title{Environmental Monitoring for Belle~II}

\author{\IEEEauthorblockN{Seokhee Park, Youngjoon Kwon}
\IEEEauthorblockA{Yonsei University\\
Seoul, Korea\\
Email: seokhee.park@yonsei.ac.kr\\
yjkwon63@yonsei.ac.kr}
\and
\IEEEauthorblockN{Mikihiko Nakao, Sadaharu Uehara}
\IEEEauthorblockA{KEK\\
Tsukuba, Ibaraki, Japan\\
Email: mikihiko.nakao@kek.jp\\
sadaharu.uehara@kek.jp}
\and
\IEEEauthorblockN{Tomoyuki Konno}
\IEEEauthorblockA{Kitasato University\\
Tokyo, Japan\\
Email: tkonno@kitasato-u.ac.jp}
}


%Telephone: (800) 555--1212, Fax: (888) 555--1212}


\maketitle

\begin{abstract}
The Belle~II experiment has just started, searching for physics beyond the
Standard Model in $B$, charm and $\tau$ decays using data with the integrated
luminosity goal of $50 ~\mathrm{ab}^{-1}$.  Before the physics run with full
detector system being installed, Belle~II Phase 2 run is on-going at the time
of the conference, until July 2018.  In this presentation, we describe the
environmental monitoring system with an emphasis on the software tools to help
the experts and the non-expert shifters who operate the experiment.  The
monitoring tools are prepared on the control room especially for the
shift-takers.  It consists of thre components: the monitoring GUI, the alarm
system, and the archiver.  The monitoring GUI shows the current state of the
detector and the alarm system generate warning states from monitored variables
with sound and email notification.  The archiver is collecting data on single
server and provide collected data to the experimental collaborators.
\end{abstract}


% \fi
\IEEEpeerreviewmaketitle

\section{Introduction}

To search for physics beyond the Standard Model and make precision measurements
of CP violation phenomena, the Belle~II experiment using the SuperKEKB collider
at KEK, Tsukuba, Japan, started taking data in April, 2018. It aims for
collecting a huge number of $B$ and charm meson and $\tau$ lepton decays, as a
successor of the Belle experiment that discovered CP violation in $B$ meson
decays.

The final goal of Belle~II\cite{Abe:2010gxa} integrated luminosity is to
accumulate $50 ~\mathrm{ab}^{-1}$ by the year 2025.  Belle~II and SuperKEKB are
going through the commissioning stage of three phases: the Phase 1 with no
Belle~II detector for the acceleration commissioning which was successfully
finished, the Phase 2 with the Belle~II detector except for the vertex detector
which is being pursued, and then the final Phase 3 run with the full detector.
During the Phase 2 run period, Belle~II plans to collect integrated luminosity
of around $50 ~\mathrm{fb}^{-1}$ until July 2018.

The Belle~II detector consists of 6 subdetector systems, for which the readout
systems are installed inside and around the detector.  Backend readout systems
reside in the Electronics Hut (Ehut) located next to the detector, where various
monitoring systems are also placed.  We use two backbone systems and one
graphical user interface (GUI) tool to distribute and share the monitoring
information: a custom-made Network Share Memory (NSM)\cite{Nakao:nsm} tools,
which has been used in Belle and was updated for Belle~II, EPICS\cite{epics},
which is widely used in high energy physics experiments, and Control System
Studio (CSS)\cite{css}, with native EPICS handling and an additional plugin for
NSM.

The Belle~II control room is located in the underground side room of the
experiment hall where the Belle detector is placed.  The Belle~II detector is
operated by two control room shifters.  In addition, a safety shifter is
assigned for the safety of the experiment, and subdetector shifters take care of
their subdetector systems.  Our environmental monitoring tools provide the
information of the detector to these shifters and store data on the central DAQ
database server.

\begin{figure}[!b] 
\centering
\includegraphics[width=3.5in]{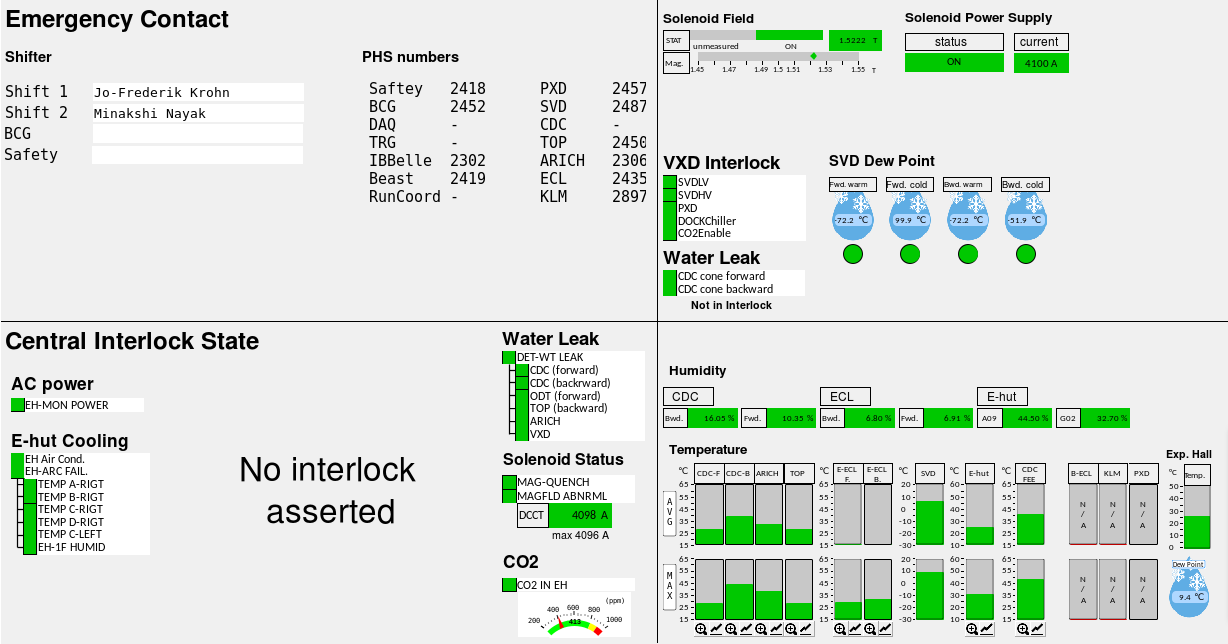}
\caption{Monitoring GUI installed in control room} 
\label{park1} 
\end{figure}

\section{Monitoring GUI}

The monitoring GUI screens are constructed on top of the CSS platform, which is
chosen for the common platform of all Belle~II user interface programs including
run control and slow control.  These tools are used not only by the detector
experts, but also for the shifters in the control room, who are most likely
non-experts on the system.

The role of the monitoring GUI is to provide comprehensive information of the
environmental status of the Belle~II detector including as many variables yet in
a visible size from a distance in a single screen.  The nominal location is on a
50-inch 4K resolution monitor on the wall of the control panel.  We consider the
following points:

\begin{enumerate} 
\item The GUI contains summarized information, e.g.  average
  and maximum values for temperatures.  In addition, small buttons are
  provided to display the detail information.  
\item Problems can be easily
  noticed and identified from distance. For this purpose, a unified color
  scheme is used. Normal, warning, and fatal states use green, orange, and red
  color respectively.  The sizes of font and widget are well optimized to fit
  the control room display.  
\item While it can be opened on a smaller 4K
  monitors in front of shifters or other users.  The panel is divided into
  $2 \times 2$ regions in such a way that each region can also fit a more
  commonly used FullHD resolution.  
\end{enumerate}

\begin{figure}[!tbh]
\centering
\includegraphics[width=3.5in]{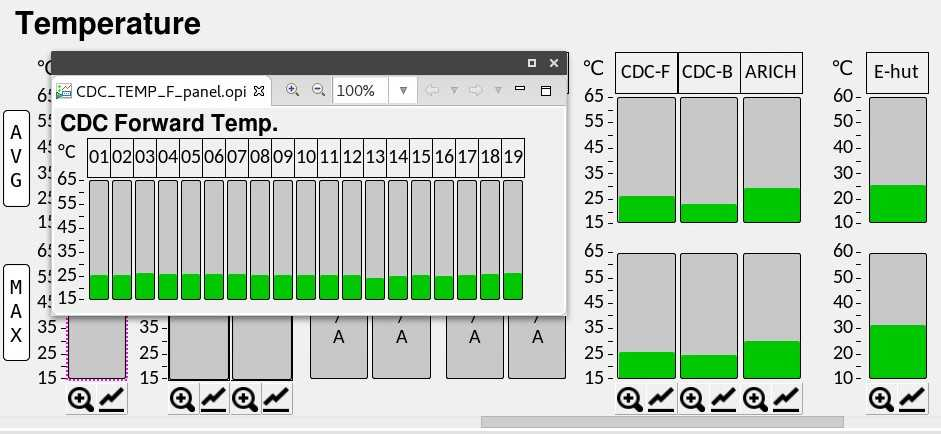}
\caption{Temperature widgets and related detail information} 
\label{park2}
\end{figure}

\section{Archiver}

To store the detector information, provided in the form of NSM variables and
EPICS process variables (PVs), we use the EPICS Archiver
Appliance\cite{archiver}.  The archiver runs on the central DAQ database server,
to record the history of various environemental variables such as temperature,
humidity, magnetic field, voltage, luminosity, and so on.  In order to archive
NSM variables, we prepared an NSM $\to$ EPICS conversion process (nsm2cad),
because the EPICS Archiver Appliance can record EPICS PVs only.  At this moment,
the number of actively archived variables is 2377.

The management system (mgmt webapp) is protected by ID and password, to prevent
modifications by unauthorized users.  The viewer is open to anyone who can
access the DAQ network.  People can see the trend graphs through a web viewer or
CSS.  The archiver is also connected to the monitoring GUI.  Button sets (small
graph button in Fig. \ref{park2}) are prepared so that people can easily find
the trends of monitoring variables.

\begin{figure}[!thb]
\centering
\includegraphics[width=3.5in]{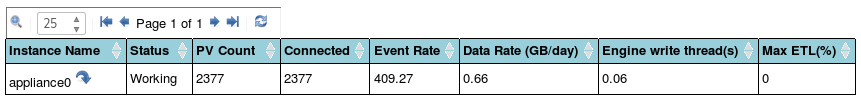}
\caption{Appliance metric}
\label{park3}
\end{figure}

\begin{figure}[!tbh]
\centering
\includegraphics[width=3.5in]{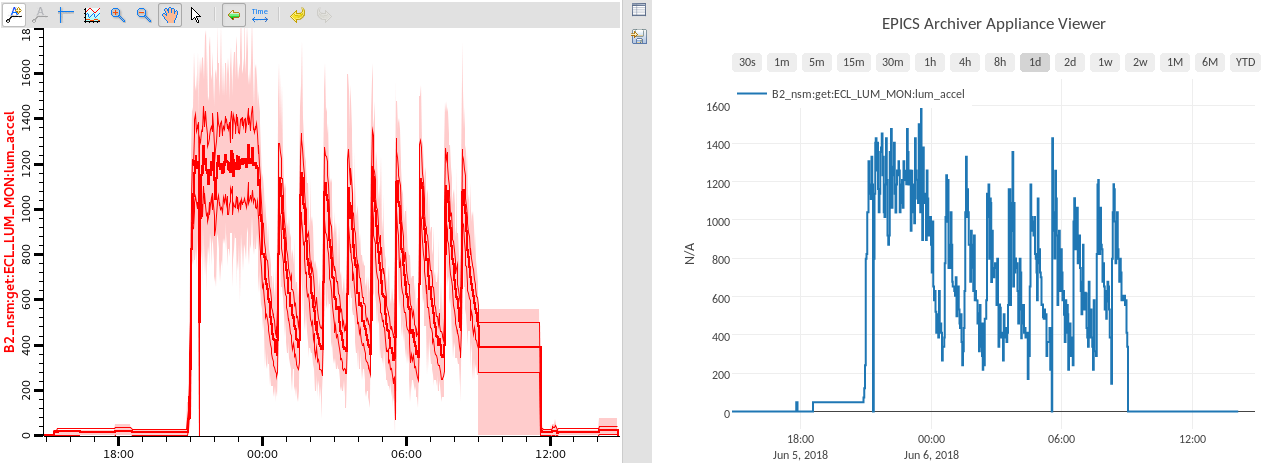}
\caption{Archiver viewer on CSS (left) and on a web browser (right)}
\label{park4} 
\end{figure}

\section{Alarm System}

The purpose of the alarm system is to provide a sound warning for control room
shifters and send email notification to subdetector experts.  The alarm system
is made of two parts, a daemon process and a GUI. The daemon is an NSM
application, and the GUI is a CSS application.

One of the important aspects is to treat enormous number of variables.  We made
the alarm daemon to collect multiple values from which a single combined warning
state is generated.  Alarm daemons also send email notifications to each
responsible expert.  The warning state is sent to the GUI, and the GUI generates
a sound for shifters.  The GUI can include and exclude each variable for sound
and email alarm. 

The alarm system is in the commissioning stage.  The connection between GUI and
daemon-GUI works fine, while we are currently developing the alarm daemon
itself.  

%For example, when collecting values from each variable, there is some
%unexpected delay in the response.

\begin{figure}[!tbh]
\centering
\includegraphics[width=3.5in]{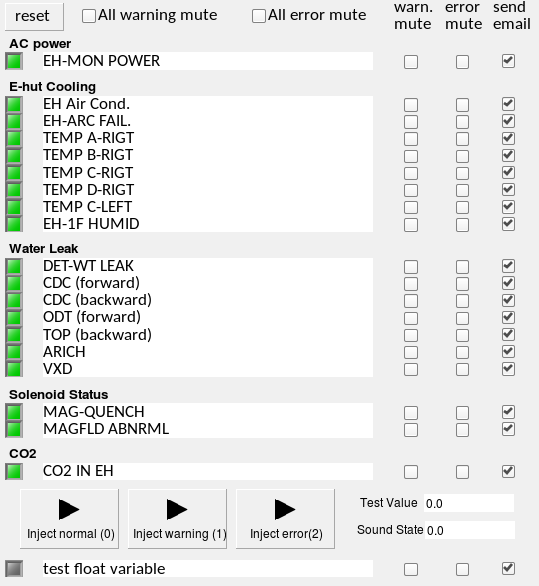}
\caption{Alarm GUI for central interlock}
\label{park5}
\end{figure}

\begin{figure}[!tbh]
\centering
\includegraphics[width=3.2in]{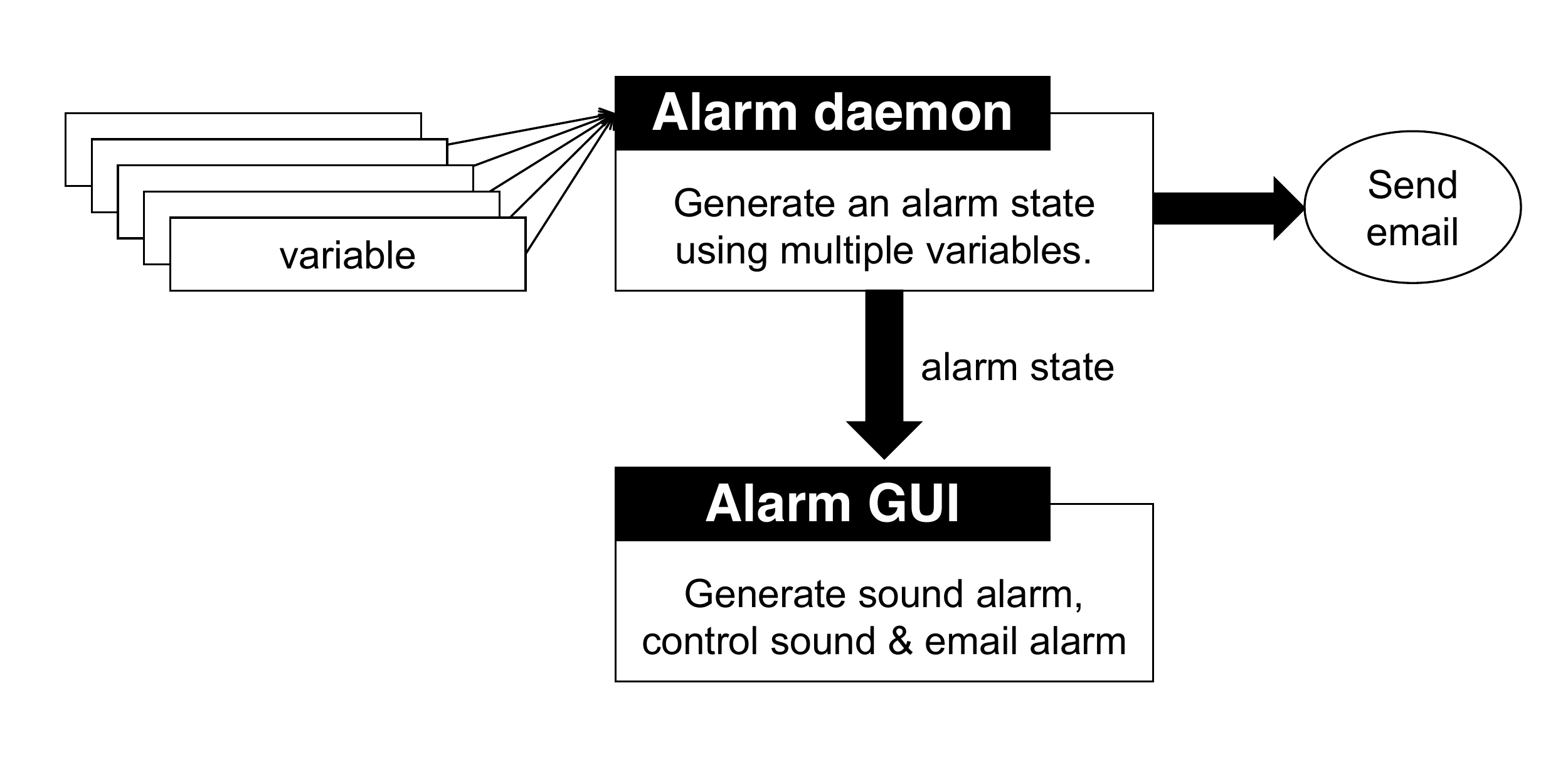}
\caption{Conceptual plot of alarm system}
\label{park6}
\end{figure}

\section{Conclusions and Plan}

The Belle~II Phase 2 operation is successfully being carried out, with the
environmental monitoring tools being one of the crucial components.  The
monitoring tools have been improved upon various experiences during the
commissioning of the Belle~II detector, while there are still remaining works to
be done.  These will be solved before the Phase 3 operation with the fully
integrated Belle~II detector.  Here are the action plans:

\begin{enumerate} 
\item {\it Monitoring GUI}
  A detection and refresh mechanism should be prepared for when the values are
  not updated, due to the frozen CSS panel or by a malfunction of the PC.

\item {\it Archiver}
  A procedure to clone the archived data should be prepared, especially for use
  in the offline analysis.  The computing resources are still available for
  additional PVs. The number of PVs can be increased to handle by an order of
  magnitude.

\newpage \IEEEtriggeratref{8}

\item {\it Alarm System}
  The commissioning of the alarm system should be completed.  The alarm system
  is not used yet for the daily operation during the Phase 2 operation.

\end{enumerate}

\end{document}